\documentclass[sigconf]{acmart}
\copyrightyear{2018}
\acmYear{2018}
\setcopyright{iw3c2w3}
\acmConference[WWW 2018]{The 2018 Web Conference}{April 23--27, 2018}{Lyon,
France}
\acmBooktitle{WWW 2018: The 2018 Web Conference, April 23--27, 2018, Lyon, France}
\acmPrice{}
\acmDOI{10.1145/3178876.3186051}
\acmISBN{978-1-4503-5639-8/18/04.}
\author{Junxiang Wang}
\affiliation{%
  \institution{George Mason University}
  \streetaddress{4400 University Drive}
  \city{Fairfax}
  \state{Virginia}
  \country{United States}
  \postcode{22030}
}
\email{jwang40@gmu.edu}

\author{Liang Zhao}
\affiliation{%
  \institution{George Mason University}
  \streetaddress{4400 University Drive}
  \city{Fairfax}
  \state{Virginia}
  \country{United States}
  \postcode{22030}
}
\email{lzhao9@gmu.edu}
\usepackage{booktabs} 
\usepackage{url}
\usepackage{algorithm}
\usepackage{algorithmic}
\usepackage{booktabs} 
\usepackage{indentfirst}
\usepackage{balance}
\usepackage{mathtools}
\usepackage{amssymb}
\usepackage{amsthm}
\usepackage{bbm}
\usepackage{bm}
\usepackage{color}
\usepackage{multirow}
\usepackage{graphicx}
\newcommand{\tabincell}[2]{\begin{tabular}{@{}#1@{}}#2\end{tabular}}
\setcopyright{rightsretained}
\begin{document}
\title{Multi-instance Domain Adaptation for Vaccine\\ Adverse Event Detection}
\begin{abstract} Detection of vaccine adverse events is crucial to the discovery and improvement of problematic vaccines. To achieve it, traditionally formal reporting systems like VAERS support accurate but delayed surveillance, while recently social media have been mined for timely but noisy observations. Utilizing the complementary strengths of these two domains to boost the detection performance looks good but cannot be effectively achieved by existing methods due to significant differences between their data characteristics, including: 1) formal language v.s. informal language, 2) single-message per user v.s. multi-messages per user, and 3) one class v.s. binary class. In this paper, we propose a novel generic framework named Multi-instance Domain Adaptation (MIDA) to maximize the synergy between these two domains in the vaccine adverse event detection task for social media users. Specifically, we propose a generalized Maximum Mean Discrepancy (MMD) criterion to measure the semantic distances between the heterogeneous messages from these two domains in their shared latent semantic space. Then these message-level generalized MMD distances are synthesized by newly proposed mixed instance kernels to user-level distances. We finally minimize the distances between the samples of the partially-matched classes from these two domains. In order to solve the non-convex optimization problem, an efficient Alternating Direction Method of Multipliers (ADMM) based algorithm combined with the Convex-Concave Procedure (CCP) is developed to optimize parameters accurately. Extensive experiments demonstrated that our model outperformed the baselines by a large margin under six metrics. Case studies showed that formal reports and extracted adverse-relevant tweets by MIDA shared a similarity of keyword and description patterns.
\end{abstract}
%
%
\maketitle

\section{Introduction}
\indent In recent decades, information extraction from social media data such as Twitter data have demonstrated to be successful in the healthcare applications \cite{lamb2013separating} \cite{iso2016forecasting} \cite{paul2011you}. Compared with existing adverse event reporting systems, social media have the following advantages: \textbf{(1). Message timeliness:} different from deliberate check from health experts, which may take months to release reports, messages regarding symptom descriptions of vaccine adverse events can be posted immediately by portable mobile devices \cite{zhao2016online}. \textbf{(2). Sensor 
ubiquity:} social media can capture ubiquitous disease information from
social sensors because they reflect the  mood and trend of the public, which can be utilized to detect vaccine adverse events. However, social media still suffer from two challenges: \textbf{1. Prohibitive labeling efforts:} in order to obtain accurate labels, it is mandatory to check all messages of all users. For example, suppose a user has 100 messages on average, labeling 1,000 users amounts to checking 100,000 messages, which can not be completed manually. \textbf{2. Class imbalance:} in practice, the proportion of positive users, whose messages indicate adverse events, is very low. As a result, a classifier is biased towards negative users,  causing high false negative rates.\\
\indent Therefore, formal reports are accurate but with poor timeliness while social media are timely but more imbalanced and labor-intensive to label. To overcome their respective drawbacks, we innovatively propose to integrate their complementary strengths. However, the integration of these two domains is seriously challenged by several salient differences between their characteristics: \textbf{1. Formal language versus informal language.} Generally, word usage in the formal reports and social media are totally different: health experts or doctors tend to use formal words or terminologies in the formal reports whereas informal words are common in the social media messages. Table \ref{tab: tweet example} gives two examples of formal reports and tweets, respectively. Keywords are shown in bold types. Medical terminologies like 'parotid', 'gland' and 'malaise' are frequently used in the formal reports while social media users tend to use informal words like 'damn' and 'Ouch'. Even some commonly used keywords both in the formal reports and social media messages like 'headache' and 'sore' differ in word frequencies. \textbf{2. Single text versus multiple messages.} Formal reports and social media also differ in structures: each reporter typically write only one report in formal reporting systems whereas each social media user can publish thousands of postings. The first Twitter example shown in Table \ref{tab: tweet example} indicates that this user has multiple tweets, whereas every symptom text belongs to only one formal report. \textbf{3. Binary class versus one class.} Typically, social media users consist of a small portion of positive users and a majority of negative ones, whereas formal reports only belong to the positive class. As is illustrated in Table \ref{tab: tweet example}, the first and the second Twitter user belong to the negative and the positive class, respectively.\\ 
\begin{table}
\centering
\small
 \caption{Two formal report and tweet examples, respectively: (+) stands for positive formal reports or tweets, (-) denotes negative tweets. Keywords are shown in bold types.}
 \begin{tabular}{c|c}
 \hline\hline
 Formal report&Tweet\\ \hline
 \tabincell{c}{The patient started to \\feel an \textbf{itching} feeling.(+)}&\tabincell{c}{1. Flu shots in Town Lobby from 1-5 pm.(-)\\
2. a flu shot for only 12 dollars.(-)}\\ \hline
 \tabincell{c}{Swollen \textbf{parotid glands},\\fever, headache, \textbf{malaise}.(+)}&\tabincell{c}{ 1.\textbf{Ouch}! so sore my arms are!\\ \textbf{damn} flu shot!(+)}\\\hline\hline
 \end{tabular}
 \label{tab: tweet example}
 \end{table}
\indent In order to simultaneously deal with these challenges, we propose a novel Multi-instance Domain Adaptation (MIDA) framework for vaccine adverse event detection by maximizing the synergy of formal reporting systems and social media data such as Twitter data. Specifically, given commonly used keywords both in formal reports and social media messages (e.g., tweets), a generalized MMD-based \cite{long2013transfer} criterion is proposed to measure the difference between the heterogeneous messages from these two domains. These message-level distances are then synthesized to user-level by a novel mixed instance kernels induced by a \emph{max rule}. Finally, a \emph{partial class-matching strategy} is leveraged to optimize the seamless integration of the two domains with different number of classes for accurate adverse event detection. The parameter optimization of MIDA is a nonconvex problem, an Alternating Direction Method of Multipliers (ADMM) \cite{boyd2011distributed} based algorithm combined with the Convex-Concave Procedure \cite{lipp2016variations} has been developed to optimize variables in a distributive manner. One real vaccine adverse event detection dataset  demonstrated that the MIDA outperformed all the baselines.\\
\indent The main contributions of our research are summarized as follows:
\begin{itemize}
\item \textbf{Design a generic framework MIDA for cross-domain adverse event detection.} The adverse event detection techniques from formal reporting systems and social media mining focus on different but complementary aspects. Since their advantages complement with each other, MIDA is proposed to integrate the strengths of them to achieve a synergy.
\item \textbf{Propose new models for multi-instance domain adaptation.} To model the word frequency differences between formal reports and social media data such as tweets, we propose a generalized MMD-based criterion and new kernels induced by the max rule in the multi-instance learning setting. 
\item \textbf{Develop an efficient nonconvex optimization algorithm.} The optimization problem is non-convex due to the introduction of the generalized MMD. An effective approach based on ADMM is developed to optimize it, where the non-convex subproblem is efficiently solved by sufficiently exploring its convex-concave property using CCP \cite{lipp2016variations}, which ensures local convergence.
\item \textbf{Conduct extensive experiments for performance evaluations.} The results on the real-world adverse event dataset demonstrate that MIDA consistently dominated the performance. Sensitivity analysis and scalability analysis on several factors are discussed thoroughly. Case studies show that formal reports and extracted adverse-relevant tweets by MIDA shared a similarity of keyword and description patterns.\end{itemize}
\indent The rest of the paper is organized as follows. In Section \ref{sec:related work}, we summarize recent research work related to this paper. In Section \ref{sec:problem setup}, we present the problem formulation. In Section \ref{sec:MIDA}, we propose the novel MIDA framework. In Section \ref{sec:optimization}, we develop an effective ADMM-based optimization algorithm. In Section \ref{sec:experiment}, extensive experiments are conducted to validate the effectiveness of our model. Section \ref{sec:conclusion} concludes by summarizing the whole paper.
\section{Related Work}
\label{sec:related work}
\indent This section introduces related work in several research fields.\\
\textbf{Multi-instance learning.} Multi-instance learning is a variant of traditional machine learning methods in which a data point is presented as a bag of multiple instances. Multi-instance classifiers are categorized as either \textit{instance-level} or \textit{bag-level} \cite{Amores2013Multiple}. \textit{Instance-level} classifiers score each instance without considering the characteristic of the whole bag. For example, the image classification of beaches and non-beaches is determined by their visual content \cite{Amores2013Multiple}; Kumar and Raj detected audio events based on a collection of audio
recordings \cite{Kumar2016Audio}.
\textit{Bag-level} is more common than \textit{instance-level}. For example, Dietterich et al. evaluated a drug as effective if it binded well with  target binding sites \cite{Amores2013Multiple} \cite{Dietterich1997Solving}.
Andrews et al. gave instance-level and bag-level formulations as a maximum margin problem in their Support Vector Machines (SVM) settings \cite{Andrews2002Support}. Zhou et al. developed two methods to discriminate bag labels with graph theory \cite{Zhou2008Multi}. However, to the best of our knowledge, very little work has applied multi-instance learning frameworks to social media applications. \\
\textbf{Transfer learning.} The idea of transfer learning lies in learning objects in the target domain with the help of knowledge transfer from the source domain \cite{pan2010survey} \cite{weiss2016survey}. Typically, transfer learning approaches are categoried as either homogeneous or heterogeneous. In homogeneous models, the source and the target share the same domain space, but probability distributions are totally different \cite{weiss2016survey}. For example, Daume III proposed an easy domain adaptation approach by feature augmentation \cite{daume2009frustratingly}. Pan et al. aligned domain-specific words into unified clusters with the help of domain-independent words for sentiment analysis \cite{pan2010cross}. Chattopadhyay et al. presented a novel framework that minimized conditional probability distribution differences between multiple subjects \cite{chattopadhyay2012multisource}. For heterogeneous models, the source and the target are represented by different feature spaces. For example,   Duan et al. projected the source and the target spaces into a common subspace and then two mapping functions were proposed to augment features \cite{duan2012learning}. Kulis et al. transferred object models from the source to the target by a nonlinear transformation \cite{kulis2011you}. Zhu et al. enriched the representation of targeted images with semantic concepts extracted from annotated source images by a matrix factorization approach \cite{zhu2011heterogeneous}. Most transfer learning work focused on the single instance, only several papers considered transfer learning in the multi-instance learning setting: Zhang and Si proposed a novel method where the target classifier was the linear combination of multiple source classifiers \cite{zhang2009multiple}; Wang et al. transferred cross-category knowledge to boost the target learning task, and a data-dependent  mixture model was presented to combine a weak classifier with multiple source classifiers \cite{wang2014adaptive}; Wang et al. mapped a target multi-instance bag into a bag-level feature space by a domain transfer dictionary, then a linear adaptive function was applied to a bag-level feature vector \cite{wang2016domain}. However, none of them focused on distance minimization between two domains.\\
\textbf{Adverse event surveillance and detection.} Recently, healthcare topics on social media have begun to attract considerable attention of researchers. Flu surveillance is an important application to mention. For instance, Lee et al. detected seasonal flu by a real-time analysis of Twitter data \cite{K2013Real}; Chen et al. made inference about a user's hidden state according to his tweets during flu outbreaks, and state statistics were aggregated by geographic region \cite{Chen2014Flu}. Signorini et al. kept track of H1N1 flu and measured flu activities \cite{signorini2011use}, while Lampos et al. utilized a Twitter microblogging service to track the flu-like illness in the United Kingdom\cite{lampos2010flu}. 
Drug-related adverse event detection is another popular application. For example, Metke et al. discussed the effect of the text-processing step on the drug adverse event detection performance \cite{Metke2014Evaluation}. Yomtov and Gabrilovich aggregated search log of Internet users to extract drug-related   adverse reactions \cite{Yomtov2013Postmarket}. However, very little work has discussed the application of vaccine adverse event detection on social media.
\section{Problem Setup}
\label{sec:problem setup}
\indent In this section, the problem addressed by this research is formulated in the formal reports as the source domain and the Twitter messages as the target domain. Section \ref{sec:problem formulation} defines the problem of vaccine adverse event detection; and Section \ref{sec:challenges} discusses challenges of the problem. 
 \subsection{Problem Formulation}
 \label{sec:problem formulation}
 \begin{table}\small
 \centering
 \caption{Important notations and descriptions}
 \begin{tabular}{cc}
 \hline
 Notations&Descriptions\\ \hline
 $X_u$& The input matrix from user $u$\\
 $Y_u$& The predefined label from user $u$\\
 $K$& The common keyword set\\
 $U$& The user set\\
 $R$& The formal report set\\
 $\beta$& The coefficient vector of  the keyword set.\\
 $c$& The formal report and Twitter data fold.\\
 $n_u$& The tweet number from user $u$.\\
 $U_p$& The positive user set.\\
 $I(u)$& An index set from user $u$.\\
\hline
 \end{tabular}
 \label{tab:notations}
 \end{table}
\indent The problem formulation addressed by this paper is given in this section. Table \ref{tab:notations} displays important notations and descriptions. Formal reports and Twitter messages are considered as the source and the target domain, respectively. $K$ denotes a common keyword set that represents symptom descriptions of vaccine adverse events in both domains, and $R$ denotes the formal report set. The $j$th entry of the $i$th formal report $R_i$, denoted by $R_{i,j}$, is the count of the $j$th keyword in the $i$th formal report $R_i$. $r$ is the number of formal reports. A tweet set is denoted as $D=\{D_u\}_{u\in U}$, where a user set is denoted as $U$ and the matrix $D_u\in \mathbb{Z}^{n_u \times \vert K\vert}$ denotes tweets from user $u$. $n_u$ refers to the number of tweets from user $u$. $D_{u,i}$ stands for the $i$th tweet from user $u$. The $j$th entry of $D_{u,i}$, denoted by $D_{u,i,j}$, is the count of the $j$th keyword in the $i$th tweet from user $u$. A user set is denoted as $U$. $Y^R=1$ represents health states indicated by formal reports, which belong to the positive class. $Y_u \in \{0,1\}$ denotes the health state of user $u$, $Y_u=1$ implies that user $u$ is regarded as a positive user (i.e. this user suffers from vaccine adverse events),  while $Y_u=0$ shows that user $u$ is negative (i.e. this user receives safe vaccines). $Y=\{Y_u\}_{u \in U}$ denotes the health states of all users. The input matrix for user $u$ is defined as $X_u=[\bm{1}_{n_u \times 1},D_u]$ where $\bm{1}_{n_u \times 1}$ is an all one vector. The dimension of $X_u$ is $n_u\times(\vert K\vert +1)$. $X_{u,i}$ denotes the $i$th row of $X_u$. $X=\{X_u\}_{u \in U}$ denotes the input matrices of all users.  Then vaccine adverse event detection problem can be formulated as follows:\\
\textbf{Problem Formulation}: Given the input matrices $X=\{X_u\}_{u \in U}$ and formal reports $R$, the goal of the problem is to detect the health state of a user $u\in U$ by learning the mapping $f$:
\begin{align}
f:\{ X_{u,1},X_{u,2},\cdots,X_{u,n_u} |R\} \rightarrow Y_u
\label{eq:problem}
\end{align}
\subsection{Challenges}
\label{sec:challenges}
\indent In order to solve the vaccine adverse event detection problem \eqref{eq:problem}, we still need to tackle several challenges.\textit{1) Distribution gap.} The formal report set $R$ and the tweet set $D$ share the same keyword space, but they differ in linguistic form and word frequency. \textit{2) Structure difference.} According to the problem formulation, the $i$th formal report $R_i$ is encoded by a vector, whereas the tweet set $D_u$ is represented by $n_u$ vectors. Different structures make distance measurement very difficult. \textit{3) Class-pattern inconsistency.} All formal reports have predefined labels $Y^R=1$, while a Twitter user $u$ is labeled as $Y_u$ where $Y_u=\{0,1\}$ has two possibilities. Thus in the next section, a novel multi-instance transfer learning model is proposed to address these problems in turn. 
\begin{figure}
\center
\includegraphics[width=0.9\columnwidth]{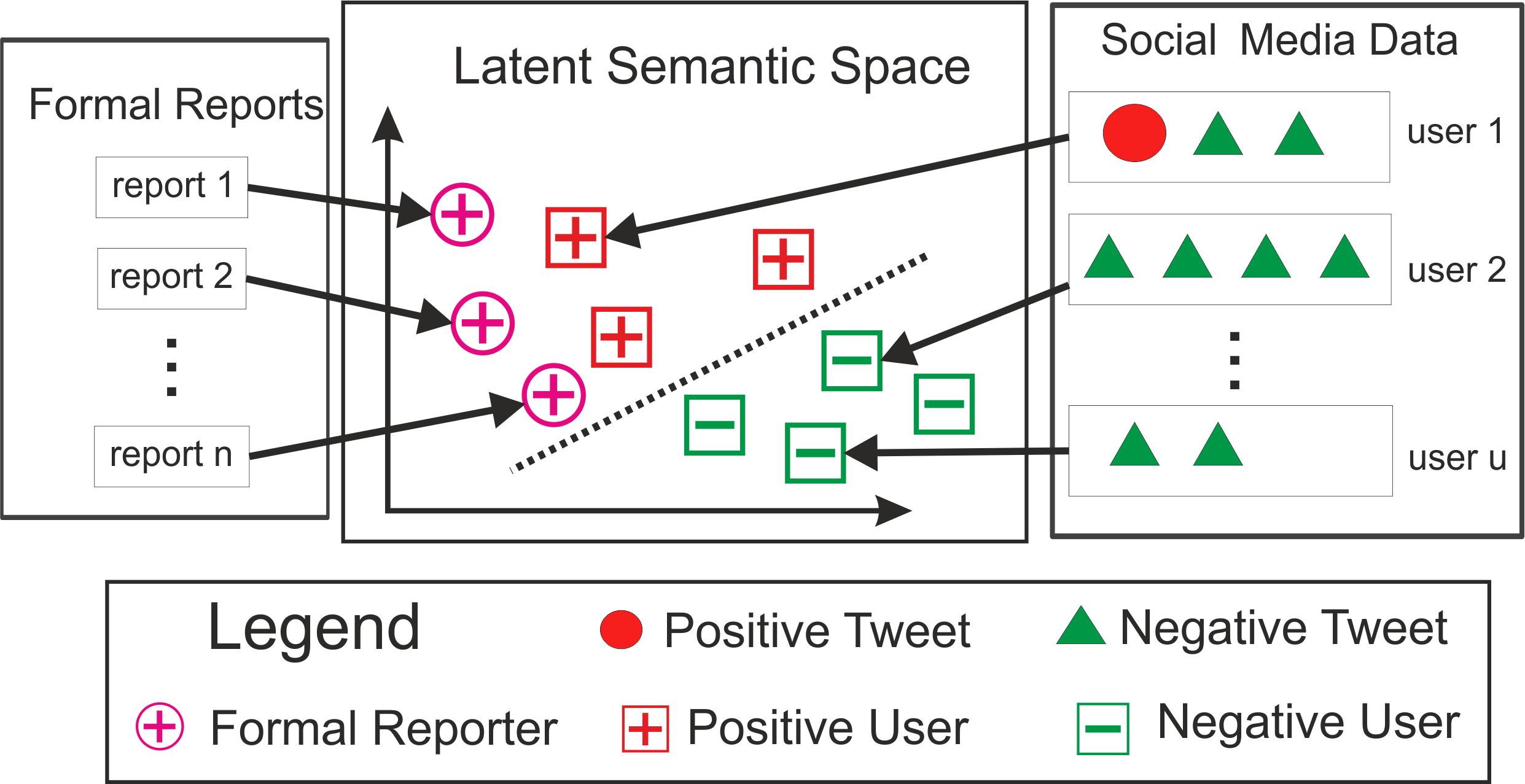}
\caption{The framework overview: combine Twitter data with formal reports to detect vaccine adverse events.}
\label{fig:framework overview}
\end{figure}
\section{Multi-instance Domain Adaptation (MIDA) Model}
\label{sec:MIDA}
\indent In this section, we develop the novel MIDA model. Specifically, a simple but effective max rule and a multi-instance classifier are discussed in Section \ref{sec:max rule}; Section \ref{sec:Domain adaptation} serves to minimize distances between formal reports and tweets; Section \ref{sec:overall model} gives the complete framework of our model, discusses several computational issues and shows the  relationship between our model and several previous methods.
\subsection{The Max Rule and Multi-instance Classifier}
\label{sec:max rule}
\indent We begin by establishing the mapping from social media users to their postings (e.g., tweets). Based on all the tweets of each user, the user is labeled as positive if at least one tweet is positive. Otherwise, that user is classified as negative. Suppose $p_{u,i}$ denotes the probability that the $i$th tweet of user $u$ is indicative of vaccine adverse events (i.e., positive). Based on the above intuition, the probability $p_u$ that the user herself is positive, is calculated by the following \textbf{max rule}:
\begin{align}
p_u=max_{i=1,\cdots,n_u}p_{u,i}
\label{eq:max rule}
\end{align}
\indent As is displayed in Figure \eqref{fig:framework overview}, positive users are composed of at least one positive tweet, which are denoted by red circles and green triangles. The max rule assigns the first Twitter user in the right of Figure \ref{fig:framework overview} a positive label. The max rule leads to asymmetric property from users to tweets because a positive user entails only a tweet indicating adverse events. \\
\indent We choose a logistic regression classifier because of its probability output. Suppose  $\beta$ is a coefficient vector where the $i-th$ element $\beta_i$ denotes the weight of the $i-th$ keyword from the keyword set $K$, then $p_{u,i}$ is represented by the following equation.
\begin{align*}
&p_{u,i}=logit(X_{u,i},\beta)
\end{align*}
where $logit(\bullet)$ is a logit function. Since our ultimate goal is to learn a model in the Twitter domain, we adopt the empirical risk minimization principle\cite{vapnik1998statistical} and the log loss function for user $u$ $Loss_u(\beta)$ is given by the following equation.
\begin{align}
Loss_u(\beta)=-Y_u\log(p_u)-(1-Y_u)\log(1-p_u)
\label{eq:log loss}
\end{align}
\subsection{Heterogeneous Domains Adaption}
\label{sec:Domain adaptation}
\indent In order to achieve the seamless integration of the knowledge from formal reports and social media, heterogeneity between these two domains needs to be considered and addressed. As shown in Figure \ref{fig:framework overview}, their heterogeneity comes from three aspects: 1) formal reports and social media messages have different linguistic forms, denoted by circles and squares; 2) formal reports only have positive samples; and 3) each social media user has multiple instances (i.e., tweets) while each reporter only has a single instance (i.e., a formal report). To overcome the first two aspects, we propose a novel \textbf{latent-space marginal distance measurement}, while to overcome the third, we propose  \textbf{mixed instance kernels}. The details are as follows.

\noindent\textbf{1. Latent-Space Marginal Distance Measurement.} As shown in Figure \ref{fig:framework overview}, the closer the positive Twitter users are to reporters, the clearer the decision boundary is. Suppose $U_p$ denotes positive Twitter user set, we aim to minimize the distance $Dist^2(R,U_p;\beta)$ between formal reports and tweets of positive users. However, existing distance measurements such as the well-known  non-parametric criterion Maximum Mean Discrepancy (MMD) \cite{long2014adaptation,duan2012domain,long2013transfer} can not handle our problem (that first two aspects of the heterogeneity we mentioned) because it is not applicable to  two domains with different numbers of classes and multiple instances. Therefore, a generalized MMD-based measurement has been proposed, which compares the distance of only positive samples in the both source and target domain in a reproducing kernel Hilbert space (RKHS) $H$,
\begin{align*}
&\min Dist^2(R,U_p;\beta)=\min \Vert\phi(R)/r-\phi(U_p)/n_p \Vert^2_H
\end{align*}
where $n_p=\vert U_p\vert$ denotes the number of positive users and $\phi(\bullet): R\cup U_p\rightarrow H$ is a feature mapping. Suppose $Ker(\bullet,\bullet)$ is a kernel function induced by $\phi(\bullet)$ such that $\phi(x)^T \phi(y)=Ker(x,y)$, the generalized MMD can be transformed into
\begin{align}
\nonumber &\min \Vert\phi(R)/r-\phi(U_p)/n_p \Vert^2_H\\\nonumber 
=&\min \phi(R)^T\phi(R)/r^2-2\phi(R)^T\phi(U_p)/(r\times n_p)+\phi(U_p)^T\phi(U_p)/n_p^2\\=&\min \!Ker\!(R,\!R)\!/r^2\!-\!2Ker(R,\!U_p;\!\beta)/(r\!\times\!n_p)\!+\!Ker(U_p,\!U_p;\!\beta)\!/\!n_p^2
\label{eq:MMD}
\end{align}
where $Ker(R,R)$, $Ker(R,U_p;\beta)$ and $Ker(U_p,U_p;\beta)$ denote the kernel inside  $R$, the kernel between $R$ and $U_p$ and the kernel inside $U_p$, respectively. Considering $Ker(R,R)$ as constant and $Ker(R,U_p;\beta)$ and $Ker(R,U_p;\beta)$ as similarity measurements dependent on $\beta$ which will be defined in the later section, Equation \eqref{eq:MMD} shows that  minimizing the generalized MMD is equivalent of putting double weights on cross-domain similarity maximization  $Ker(R,U_p;\beta)$ at the cost of similarity minimization inside the Twitter domain $Ker(U_p,U_p;\beta)$.\\
\begin{figure}
\center
\includegraphics[scale=0.4]{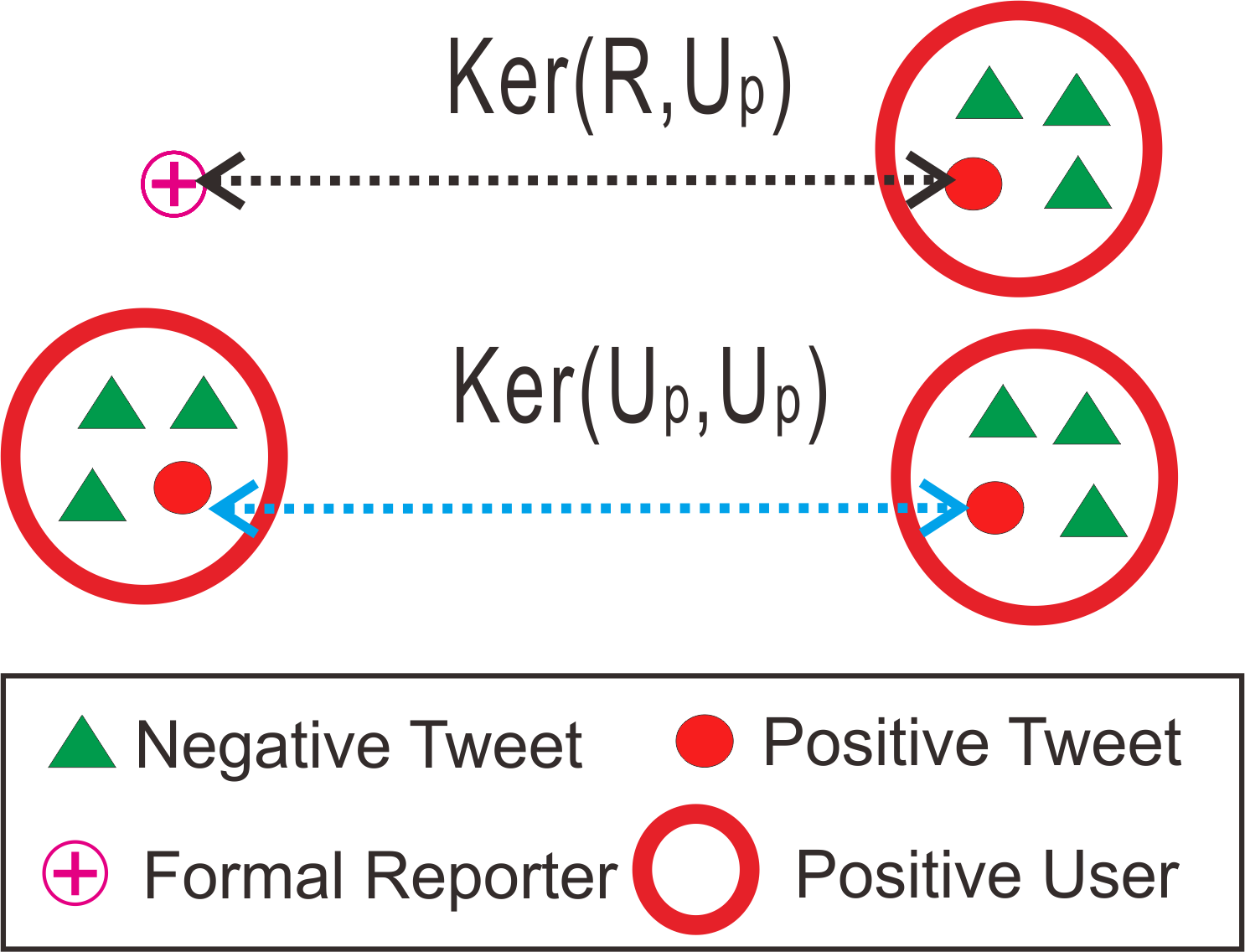}
\caption{The motivation of \emph{mixed instance kernels}: $Ker(R,U_p;\beta)$, denoted by the black double-headed arrow, is encoded by the  similarity between formal reports and positive tweets; $Ker(U_p,U_p;\beta)$, denoted by the blue double-headed arrow, is encoded by the similarity  between two positive tweets.}
\label{fig:kernel motivation}
\vspace{-0.8cm}
\end{figure}
\noindent\textbf{2. Mixed Instance Kernels.} Measuring the above distances between users and reporters (i.e., bags) require the characterization of their messages (i.e., instances), because the polarity of a user/reporter is determined by that of her messages. Mathematically, we need to determine  kernels $Ker(R,U_p;\beta)$ and $Ker(U_p,U_p;\beta)$ in Equation \eqref{eq:MMD}. As shown in Figure \ref{fig:framework overview}, the polarity of a Twitter user is collectively determined by all her tweets (i.e., the multi-instance case) while the polarity of a reporter is determined by her one and only formal report (i.e., the single-instance case). However, there is no kernel to handle this hybrid of multi-instance and single-instance inputs. This motivates us to propose novel \emph{mixed instance kernels}: as illustrated by Figure \ref{fig:kernel motivation}, kernel $Ker(R,U_p;\beta)$, denoted by the black double-headed arrow, is encoded by the similarity between formal reports and positive tweets determined by the max rule defined in Equation \eqref{eq:max rule}; kernel $Ker(U_p,U_p;\beta)$, denoted by the blue double-headed arrow, is encoded by the similarity  between two positive tweets determined by the max rule defined in Equation \eqref{eq:max rule} as well. Therefore, kernels $Ker(R,U_p;\beta)$ and $Ker(U_p,U_p;\beta)$ map user-level (i.e., bag-level) similarity measurements into message-level (i.e., instance-level) ones. Notations in Figure \ref{fig:kernel motivation} are formulated mathematically as follows.
\begin{align}
&Ker(R,U_p;\beta)=\sum\nolimits_{u\in U_p}\sum\nolimits_{i=1}^{r} Ker(D_{u,I(u)},R_i;\beta)\label{eq:kernel1}\\
&Ker(U_p\!,\!U_p;\!\beta)\!=\!\sum\nolimits_{u_1\in U_p}\!\sum\nolimits_{u_2\in U_p} \!Ker\!(\!D_{u_1,I(u_1)},\!D_{u_2,I(u_2)};\beta)
\label{eq:kernel2}
\end{align}
where an index set $I(u)=\arg\max_{i} p_{u,i}$ is introduced auxiliarily, which records the index of the tweet selected by the max rule.\\
\indent To reduce the complexity of the model, we introduce the triangular kernel $Ker(x,y)=-\Vert x- y\Vert^2_2$ \cite{fleuret2003scale}. 
After integrating Equation \eqref{eq:kernel1} and \eqref{eq:kernel2} into Equation \eqref{eq:MMD}, we have
\begin{align}
\nonumber &\min Dist^2(R,U_p;\beta)=\min 2\sum\nolimits_{u\in U_p}\sum\nolimits_{i=1}^{r} \Vert D_{u,I(u)} \!-\!R_i\Vert^2_2/(r\times n_p)\\&-\sum\nolimits_{u_1\in U_p}\sum\nolimits_{u_2\in U_p} \Vert D_{u_1,I(u_1)}-D_{u_2,I(u_2)}\Vert^2/(n_p^2)
\label{eq:MMD expression}
\end{align}
where the weighted distance measure $\Vert\bullet\Vert$ is defined as $\Vert x\Vert^2_2=\sum\nolimits_{i=1}^{\vert K \vert} x_i^2\beta_{i+1}^2
$ where $\beta_{i+1}^2(i=1,\cdots,\vert K\vert)$ represents the weight of the $i$th keyword.
\subsection{Overall Model}
\label{sec:overall model}
\indent The above consideration of new distance measurement and kernels lead to a new marginal domain adaption framework that jointly minimizes the  empirical error and the difference between the two heterogeneous domains:
\begin{align}
\beta^*&=\arg\min\nolimits_\beta L(\beta)+\lambda_2 Dist^2(R,U_p;\beta)
\label{eq:optimization}
\end{align} where $L(\beta)=\sum\nolimits_{u\in U}Loss_u(\beta)+\lambda_1\Vert\beta\Vert_1$ such that $\lambda_1>0$ is a parameter for $\ell_1$ regularization due to high sparsity of the feature set and $\lambda_2>0$ is a parameter which adjusts the weight between the $L(\beta)$ and $Dist^2(R,U_p;\beta)$. $Loss_u(\beta)$ and $Dist^2(R,U_p;\beta)$ are given in Equation \eqref{eq:log loss} and \eqref{eq:MMD expression}, respectively.
\subsubsection{Computational Issues}
\ 

\textbf{Initial instance pruning.} Many positive users in the real Twitter dataset have many tweets, but most of them are irrelevant to adverse events and thus increase the complexity of the problem. Furthermore, they are similar to the tweets from negative users. Therefore it is necessary to prune them as a preprocessing step. One popular way is to build a Kernel Density Estimator (KDE) to model the distribution of the tweets from negative users \cite{ro1973pattern} \cite{fu2009instance}. The other method is to build a one-class classifier like OSVM using event-irrelevant tweets \cite{khan2009survey}.\\
\indent \textbf{Data splitting.} The other consideration is the huge computational burden of the generalized MMD with the rapid growing size of formal reports and the Twitter data.  One intuitive but effective way is to split formal reports and Twitter data. Suppose $R=\cup_{i=1}^c R^i$ and $U_p=\cup_{i=1}^c U_p^i$ are split into $c$ partitions where $c$ is the number of partition, then  $Ker(R,U_p;\beta)=\sum\nolimits_{i=1}^c Ker(R^i,U_p^i;\beta)$ and $Ker(U_p,U_p;\beta)=\sum\nolimits_{i=1}^c Ker(U_p^i,U_p^i;\beta)$.
\subsubsection{Model Generalization}
\ 

\indent Our model can be further generalized to multiple source domains, the novelty of the generalization lies in the \emph{weighted MMD scheme}. Suppose there are $m$ source formal report data, each of which is represented as $F_i(i=1,\cdots,m)$, then the MIDA model is formulated as:
\begin{align*}
&\beta^*\!=\!\arg\min\nolimits_{\beta,w_i}\sum\nolimits_{u\in U}\!Loss_u(\beta)\!+\!\lambda_1\Vert\beta\Vert_1\!+\!\lambda_2 \sum_{i=1}^m w_i Dist^2(F_i,U_p;\beta)\\
&s.t. \sum\nolimits_{i=1}^m w_i=1, w_i\geqslant 0(i=1,\cdots,m)
\end{align*}
where $w_i(i=1,\cdots,m)$ is an generalized MMD weight for the $i$-th formal report data source.
\subsubsection{Relationship to Previous Related Approaches}
\

\indent In this subsection, we show that several classic methods are special cases of our model.\\
\indent \textbf{1. Generalization of logistic regression.} Let $n_u=1$ for $u \in U$ and $R=\emptyset$.   The model then is reduced to a logistic regression with $\ell_1$-norm regularization \cite{boyd2011distributed}.
\begin{align*}
&\beta^*\!=\!\arg\min\nolimits_\beta\!\sum\nolimits_{u\in U}Loss_u(\beta) +\!\lambda_1\!\Vert \beta \Vert_1
\end{align*}
\indent \textbf{2. Generalization of logistic regression combined with transfer learning.} Let $n_u=1$ for $u\in U$. The model is then reduced to a logistic regression combined with transfer learning \cite{duan2012domain}.
\begin{align*}
&\beta^*=\arg\min\nolimits_\beta \sum\nolimits_{u\in U} Loss_u(\beta) +\lambda_1\Vert \beta \Vert_1+\lambda_2 Dist^2(R,U_p;\beta)
\end{align*}
\indent \textbf{3. Generalization of logistic regression combined with multi-instance learning.} Let $R=\emptyset$. The model is then reduced to a logistic regression combined with multi-instance learning \cite{Amores2013Multiple}.
\begin{align*}
&\beta^*\!=\!\arg\min\nolimits_{\beta}\sum\nolimits_{u\in U}Loss_u(\beta)+\!\lambda_1\!\Vert \beta \Vert_1
\end{align*}
\begin{algorithm} 
\scriptsize
\caption{the MIDA Algorithm} 
\begin{algorithmic}[1] 
\REQUIRE $X$, $Y$, $\lambda_1$, $\lambda_2$. 
\ENSURE $\beta$ 
\STATE Initialize $\beta$, $S$, $\rho$, $r=0$, $s=0$, $k=0$.
\REPEAT
\STATE Update the index set $I(u)$.
\STATE Update $\rho^{k+1}$ if necessary.
\STATE Update $S^{k+1}$ by Equation \eqref{eq:update S}.
\STATE Update $\beta^{k+1}$ by Equation \eqref{eq:update beta}.
\STATE$h^{k+1}\leftarrow h^k+\rho^{k+1}(S^{k+1}-X\beta^{k+1})$.
\STATE $r^{k+1}\leftarrow\Vert S^{k+1}-X\beta^{k+1}\Vert_2$. $\#$Calculate primal residual.
\STATE$s^{k+1}\leftarrow\Vert \rho^{k+1}X(\beta^k-\beta^{k+1})\Vert_2$. $\#$Calculate dual residual.
\STATE $k\leftarrow k+1$.\\
\UNTIL{certain convergence condition is satisfied.} 
\STATE Output $\beta$.
\end{algorithmic} 
\end{algorithm}
\section{Optimization}
\label{sec:optimization}
\indent The Equation \eqref{eq:optimization} is a non-convex and non-smooth which is very difficult to be solved by traditional optimization methods. In most recent years, ever more work utilizes ADMM to solve non-convex and non-smooth problem effectively and efficiently \cite{wang2015global,hong2016convergence}. Here in order to solve Equation \eqref{eq:optimization}, we propose a new ADMM-based algorithm. To simplify the algorithm, we introduce an auxiliary variable $S$ and reformulate the problem to its equivalence as follows:
\begin{align}
\nonumber &\beta^*=\arg\min\nolimits_{\beta} \mathop \sum\nolimits_{u \in U}(\log(1+\exp(S_{u,I(u)}))-Y_u S_{u,I(u)}) +\lambda_1 \Vert\beta\Vert_1\\\nonumber
&\!-\!\lambda\!_2\!\sum\nolimits_{u_1\in U_p}\!\sum\nolimits_{u_2\in U_p}\!\sum\nolimits_{j=1}^{\vert K\vert} (D_{u_1,I(u_1),j}\!-\!D_{u_2,I(u_2),j})^2\!\beta_{j+1}\!^2\!/\!(n_p^2)\\ &+2\lambda_2\sum\nolimits_{i=1}^{r}\sum\nolimits_{u\in U_p}\sum\nolimits_{j=1}^{\vert K\vert} (R_{i,j}-D_{u,I(u),j})^2\beta_{j+1}^2/(r\times n_p) \label{eq:reformulation}\\
\nonumber &s.t. \ S_{u,i}=X_{u,i}\beta
\end{align}
\indent The augmented Lagrangian function of Equation \eqref{eq:reformulation} is:
\begin{align*}
&L_\rho(S,\beta,h)=\sum\nolimits_{u \in U}(\log(1+\exp(S_{u,I(u)}))-Y_u S_{u,I(u)}) +\lambda_1 \Vert\beta\Vert_1\\  &\!-\!\lambda_2\!\sum\nolimits_{u_1\in U_p}\!\sum\nolimits_{u_2\in U_p}\!\sum\nolimits_{j=1}^{\vert K\vert} (D_{u_1,I(u_1),j}\!-\!D_{u_2,I(u_2),j})^2\!\beta_{j+1}^2\!/\!(n_p^2)\\&+2\lambda_2\sum\nolimits_{i=1}^{r}\sum\nolimits_{u\in U_p}\sum\nolimits_{j=1}^{\vert K\vert} (R_{i,j}-D_{u,I(u),j})^2\beta_{j+1}^2/(r \times n_p)\\&+\rho/2\Vert S_{u,i}-X_{u,i}\beta+h_{u,i}\Vert^2_2
\end{align*}
where $\rho>0$ is a penalty parameter. The MIDA algorithm is shown in Algorithm 1. Concretely, Lines 8-
9 calculate residuals and Lines 4-7 update each parameter
alternately by solving the sub-problems described below.\\
\textbf{Update $S^{k+1}$}\\
The auxiliary variable $S$ is updated as follows:
\begin{align}
\nonumber &S^{k+1} \leftarrow \arg\min\nolimits_S \sum\nolimits_{u\in U}(log(1+\exp(S_{u,I(u)}))-Y_u S_{u,I(u)})\\&+(\rho^{k+1}/2)\Vert S -X\beta^k+h^k \Vert^2_2
\label{eq:update S}
\end{align}
This subproblem is a logistic regression with an $\ell_2$ penalty term. A fast iterative shrinkage-thresholding algorithm (FISTA) \cite{beck2009fast} is applied to solve this problem because the log loss is differentiable so that each iteration has a close-form solution. \\
\textbf{Update $\beta^{k+1}$}\\
\indent The decision variable $\beta$ is updated as follows:
\begin{align}
\nonumber &\beta^{k+1} \leftarrow \arg\min\nolimits_{\beta} \lambda_1 \Vert\beta\Vert_1+\rho^{k+1}/2\Vert S^{k+1}-X\beta+h^{k}\Vert^2_2\\\nonumber 
&-\!\lambda_2\!\sum\nolimits_{u_1\in U_p}\!\sum\nolimits_{u_2\in U_p}\!\sum\nolimits_{j=1}^{\vert K\vert} (D_{u_1,I(u_1),j}\!-\!D_{u_2,I(u_2),j})^2\!\beta_{j+1}^2\!/\!(n_p^2)\\&+2\lambda_2\sum\nolimits_{i=1}^{r}\sum\nolimits_{u\in U_p}\sum\nolimits_{j=1}^{\vert K\vert} (R_{i,j}\!-\!D_{u,I(u),j})^2\!\beta_{j+1}\!^2/(r \!\times \!n_p).
\label{eq:update beta}
\end{align}
\indent Even though this subproblem is nonconvex, it can be solved by Convex-Concave Procedure (CCP), which ensures local convergence \cite{lipp2016variations}. We split this objective function further,
\begin{align*}
&l(\beta)=\lambda_1 \Vert\beta\Vert_1+\rho^{k+1}/2\Vert S^{k+1}-X\beta+h^{k}\Vert^2_2\\&+2\lambda_2\sum\nolimits_{i=1}^{r}\sum\nolimits_{u\in U_p}\sum\nolimits_{j=1}^{\vert K\vert} (R_{i,j}-D_{u,I(u),j})^2\beta_j^2/(r \times n_p)\\
&m(\beta)=\lambda_2\!\sum\nolimits_{u_1\in U_p}\!\sum\nolimits_{u_2\in U_p}\!\sum\nolimits_{j=1}^{\vert K\vert} (D_{u_1,I(u_1),j}\!-\!D_{u_2,I(u_2),j})^2\!\beta_{j+1}^2\!/\!(n_p^2).
\end{align*}  then the optimization objective becomes:
\begin{align*}
\beta^{k+1}=\arg\min\nolimits_{\beta} l(\beta)-m(\beta)
\end{align*}
The algorithm of updating $\beta$ is shown in Algorithm 2. The key idea of CCP is to convexify concave function $m(\beta)$ by linearized function $\tilde{m}(\beta)$. Now the following problem can be solved by FISTA \cite{beck2009fast} again.
\begin{align}
\beta^{q+1}=\arg\min\nolimits_{\beta} l(\beta)-\tilde {m}(\beta)
\label{eq: solve beta}
\end{align}
\indent Two important issues should be taken into account: one is to choose appropriate $\rho$ and $\lambda_2$. To guarantee the existence of the local optimum, the relationship between $\rho$ and $\lambda_2$ can be set as $\rho\geqslant10\lambda_2$ empirically. Otherwise, the CCP will lead $\beta$ to infinity. The other is the initial value of $\beta$, which affects convergence property and performance. It is recommended that an initial point of $\beta$ be chosen from the coefficient of a trained logistic regression classifier.
\begin{algorithm} 
\scriptsize
\caption{the $\beta$-update Algorithm} 
\begin{algorithmic}[1] 
\REQUIRE $S$, $X$, $\lambda_1$, $\lambda_2$, $\rho$. 
\ENSURE $\beta$ 
\STATE Initialize $\beta$, $q=0$.
\REPEAT
\STATE Convexify $\tilde{m}(\beta)\leftarrow m(\beta^q)+\nabla m(\beta^q)(\beta-\beta^q)$.
\STATE Update $\beta^{q+1}$ by solving Equation \eqref{eq: solve beta}.
\STATE $q\leftarrow q+1$.\\
\UNTIL{some convergence criterion is satisfied.} 
\STATE Output $\beta$.
\end{algorithmic} 
\end{algorithm}
\vspace{-0.3cm}
\section{Experiments}
\label{sec:experiment}
In this section, we evaluate the  MIDA using a real adverse event detection dataset, which demonstrated the effectiveness and outstanding performance of MIDA compared with existing methods. Sensitivity analysis and scalability analysis on the effect of several factors were also explored. Case studies on the formal reports and extracted adverse-relevant tweets were analyzed as well. All experiments were conducted on a 64-bit machine with Intel(R) core(TM) quad-core processor (i3-3217U CPU@ 1.80GHZ) and 4.0GB memory.
\subsection{Dataset Description}
\indent The task of the first dataset is to detect whether Twitter users are affected by adverse event according to their tweets. The dataset consists of  Twitter data and formal reports. They both were encoded by 234 keywords.\\
\indent \textbf{Input Twitter Data Retrieval.} Twitter data were analyzed in compliance with the Twitter policies\footnote{\url{https://dev.Twitter.com/overview/terms/agreement-and-policy}}. The Twitter data in this paper were retrieved by the following process. First, we queried the Twitter API to obtain the tweets that were potentially related to the topic "flu shot" by the query consisting of 113 keywords including "flu", "h1n1", "vaccine". A total of 11,993,211,616 tweets for the period between Jan 1, 2011 and Apr 15, 2015 in the United States were retrieved. Second, from the retrieved tweet sets, the Twitter users who had indicated flu vaccination were identified by their tweets using the LibShortText \cite{Yu2013LibShortText} text filter that was trained on 10,000 positive and another 10,000 negative tweets provided by Lamb et al. \cite{lamb2013separating}. The full text representations were used as the features in LibShortText. Then, we queried the Twitter API again for those users identified in the second step to obtain their tweets posted within 60 days since their vaccination were identified. The retrieved tweets formed our final Twitter data set, which contained 41,438 tweets from 1,572 users where 566 were labeled as positive users and 1,006 were negative.\\ 
\indent \textbf{Formal Reports.} We downloaded all raw data from the Vaccine Adverse Event Reporting Systems (VAERS) for the year 2016 in the Comma-separated Value (CSV) format\footnote{\url{https://vaers.hhs.gov/data/datasets.html?}}. The VAERS data consisted of 29 columns including VAERS ID, report date, sex, age and symptom text. The symptom text column contained adverse event descriptions either from patients or doctors. Each element in the symptom text column was considered as a formal report. 2500 formal reports were extracted. 
\subsection{Experimental Protocol}
\subsubsection{Parameter Settings and Metrics}
\indent We considered the MIDA for comparison. Two tuning variables $\lambda_1$ and $\lambda_2$ are included in the algorithm, which were set to 0.01 and 1 based on a five-fold cross validation on the training set, respectively. In addition, the number of partition $c$ was set to 100. The maximum number of iterations was set to 20.\\
\indent Several metrics were utilized to evaluate model performance: the Accuracy (ACC) is the ratio of accurately labeled samples to all samples; the Precision (PR) is the ratio of accurately labeled as positive samples to all labeled as positive samples; the Recall (RE) defines the ratio of accurately labeled as positive samples to all positive samples; the F-score (FS) is the harmonic mean of precision and recall; the Receiver Operating  Characteristic (ROC) curve delineates the classification ability of a model as its discrimination threshold varies; and the Area Under ROC curve (AUC) is an important measurement of classification ability; the Precision Recall (PR) curve is the other one to measure classification performance in which recall and precision are listed as the X axis and the Y axis, respectively. The Area Under PR curve (AUPR) is as important as AUC. \\
\indent Besides model performance comparison, we also explored the effect of number of iterations on the AUC and that of  number of formal reports and number of users on the running time per iteration.
\subsubsection{Comparison Methods}
\indent The following methods were utilized as baselines for the performance comparison. All
parameters in the baselines were set based on the five-fold cross validation on the training set. Baselines were categorized by either multi-instance learning methods or transfer learning methods. Method 1 and 2 belong to multi-instance learning methods, they do not need formal reports. Method 3,4 and 5 belong to the transfer learning category. For them, the input matrix $X_u$ was summed by column for user $u\in U$ on the Twitter data. \\
\indent 1. Multi-instance Learning
based on Fisher Vector representation (miFV) \cite{Wei2014Scalable}. Multiple instances were mapped into a high dimensional vector by the Fisher Vector (FV) representation. The SVM was applied to train a classifier.\\
\indent 2. Multi-instance Learning
based on the Vector of Locally Aggregated Descriptors representation (miVLAD) \cite{Wei2014Scalable}. The idea of the miVLAD was very similar to miFV, except that instances were mapped by the Vector of Locally Aggregated Descriptors (VLAD) representation.\\
\indent 3. Joint Distribution Adaptation (JDA) \cite{long2013transfer}. JDA aimed to reduce both marginal distributions and conditional distributions between the source domain and target domain. It mapped two domains into a common Hilbert space.\\
\indent 4. Graph Co-Regularized Transfer Learning (GTL) \cite{long2014transfer}. GTL complemented empirical likelihood maximization with geometric structure preservation and integrated them seamlessly into a unified framework.\\
\indent 5. Adaptation Regularization based Transfer Learning (ARTL) \cite{long2014adaptation}. The propose of ARTL was to minimize structural risk, domain distribution difference and perverse manifold consistency simultaneously.
\subsection{Performance}
In this section, experimental results for the MIDA are analyzed for all the comparison methods. Table \ref{tab:performance} summarizes  prediction results of the MIDA compared with other methods on the Twitter dataset.
\subsubsection{Model Performance on the Twitter Dataset}
\begin{figure}
\begin{minipage}{0.49\linewidth} 
\centerline{\includegraphics[width=0.9\columnwidth]{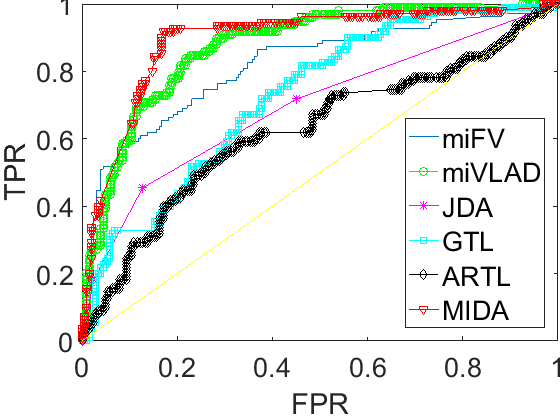}}
\centerline{(a) ROC curve}
\end{minipage}
\begin{minipage}{0.49\linewidth} 
\centerline{\includegraphics[width=0.9\columnwidth]{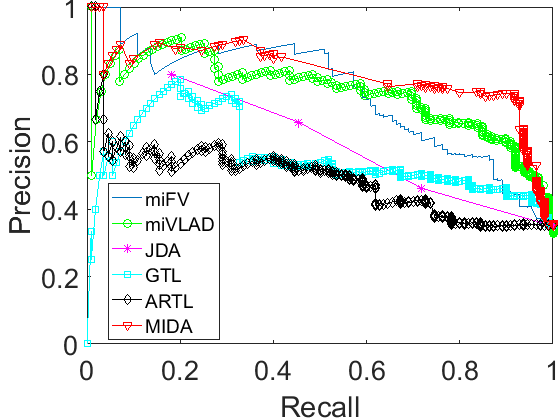}}
\centerline{(b) PR curve}
\end{minipage}
\caption{ROC curve and PR curve under the Twitter dataset: the MIDA was superior to baselines.}
\label{fig:roc}
\end{figure}

\indent  The results demonstrated in Table \ref{tab:performance} indicated that the MIDA performed better than any baseline. It ranked the first in the four metrics except the RE and the FS metric. As two most important metrics, the AUC and the AUPR, the MIDA dominated all baselines: the AUC of the MIDA was higher than $0.85$, while that of ARTL was lower than $0.6$; the AUPR of the MIDA exceeded $0.76$, whereas that of the miVLAD was only $0.70$. When it came to the ACC, the MIDA was about $0.16$ better than the GTL. The MIDA also achieved a competitive score in the RE metrics, surpassing $0.53$ whereas the PR of the JDA was only around $0.49$. As for the PR,  the MIDA performed $0.14$ better than that of the JDA. Due to excellent performance in the PR and the RE metric, the MIDA was competitive in the FS metric, which was $0.13$ better than the ARTL. The superiority of the MIDA consisted in effective utilization of formal reports by distribution matching, while multi-instance learning methods lacked formal reports and summation by column led to great information loss for transfer learning methods. Multi-instance learning methods outperformed transfer learning methods thoroughly. The ACCs of the miFV and the miVLAD were both higher than $0.76$, whereas the best score of transfer learning methods, which was achieved by the JDA, was only lower than $0.72$. The PR scores of the miFV and the miVLAD were in the vicinity of $0.71$, $0.3$ better than that of the GTL and the ARTL.\\
\indent Figure \ref{fig:roc} shows the ROC and the PR curve of the MIDA and baselines. In the ROC curve, the X axis and the Y axis denote False Positive Rate (FPR)
and True Positive Rate (TPR), respectively. In the PR curve, the X axis and the Y axis denote Recall
and Precision, respectively. Overall, the ROC curve of the MIDA covered larger area than any baselines, which was consistent with Table \ref{tab:performance}. The miFV and the miVLAD performed similarly: they both outperformed three transfer learning methods. The ARTL performed the worst of all baselines, which was slightly better than the random guess. The similar patterns were displayed in the PR curve: all baselines were surrounded by the MIDA. The PR curves of three transfer learning methods: the JDA, the GTL and the ARTL were surrounded by these of two multi-instance learning methods: the miFV and the miVLAD.
\begin{table}[!hbp]
\vspace{-0.3cm}
\small
\centering
\caption{Classification performance on the Twitter dataset under six metrics: the MIDA dominated all baselines.}
\begin{tabular}{c|c|c|c|c|c|c}
\hline\hline
Method & ACC & PR & RE & FS & AUC&AUPR \\
\hline
miFV&0.7754&0.7321&0.5965&\textbf{0.6570}&0.8451&0.7584	 	\\	\hline
miVLAD&0.7614&0.6882&\textbf{0.6245}&0.6535&0.8227&0.7053	 \\ \hline
JDA&0.7163&0.6370&0.4938&0.5552&0.7091&0.4652
\\\hline GTL&0.6158&0.4215&0.2061&0.2750&0.6310&0.4905
\\\hline
ARTL&0.5356&0.4108&0.6435&0.5003&0.5997&0.4494 
\\\hline MIDA&\textbf{0.7767}&\textbf{0.7735}&0.5333&0.6310&\textbf{0.8530}&\textbf{0.7642}\\	\hline\hline
\end{tabular}
\label{tab:performance}
\end{table}
\subsubsection{The effect of iterations on the residuals and the AUC}
\begin{figure}
\begin{minipage}
{0.49\linewidth}
\centerline{\includegraphics[width=0.9\columnwidth]
{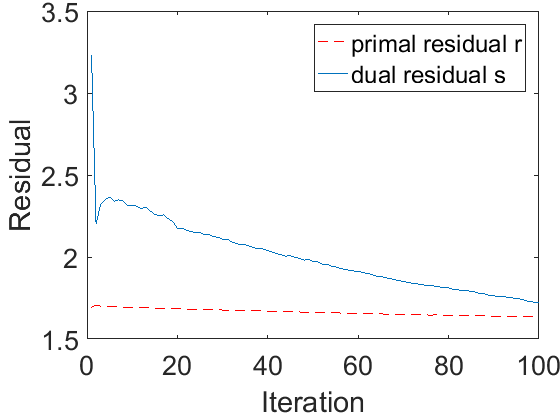}}
\centerline{(a) residual versus iteration}
\end{minipage}
\hfill
\begin{minipage}
{0.49\linewidth}
\centerline{\includegraphics[width=0.9\columnwidth]
{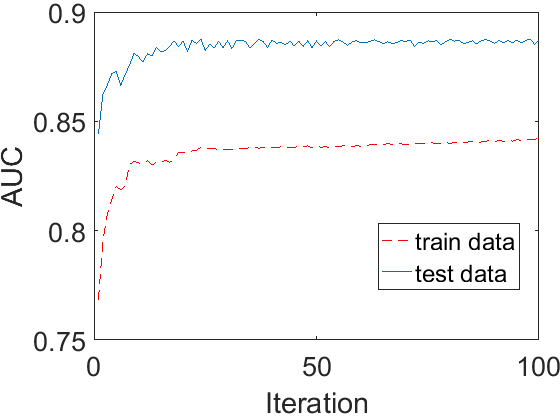}}
\centerline{(b) AUC versus iteration}
\end{minipage}
\caption{The effect of iteration on the residuals and  AUCs: $r$ and $s$ declined with iteration; the AUCs of both training  and test data increased steadily with iteration.}
\label{fig:iteration effect}
\vspace{-0.6cm}
\end{figure}

\indent We examined the effect of iterations on the residuals and the AUCs. The AUC metric was chosen because it reflected classification generalization while five other metrics were subject to change as the threshold varied.\\
\indent Figure \ref{fig:iteration effect}(a) shows the change of residuals $r$ and $s$ with respect to iteration. The primal residual $r$ remained a smooth and steady decline while the dual residual $s$ tumbled down rapidly at first, then increased slightly and finally decreased steadily to less than 2. The AUCs of training data and test data with regard to iteration are displayed in Figure \ref{fig:iteration effect}(b). Surprisingly, the AUC of test data was better than that of training data. They both increased sharply at the beginning, then the increase began to decrease as the iteration continued. This trend reflected that tens of iteration were sufficient to practical applications for the ADMM algorithm \cite{boyd2011distributed}. Another important point is that the AUCs of the training and the test data started at very high level. Therefore the ADMM algorithm achieved a satisfactory result even with several iterations.
\subsubsection{Running time per iteration}
\indent In this subsection, the relationship between running time per iteration and two potential factors, namely, the number of users, the number of formal reports was explored. The running time was calculated by the average of 20 iterations. The result is shown in Table \ref{tab:running time}. The number of users ranged from 100 to 1,500, with increasing 100 each time whereas the number of formal reports ranged from 500 to 2,500 and 500 was increased each time. Generally, the running time increased as the user set and the formal report set became larger. However, some exceptions were found in Table \ref{tab:running time}: for example, when  $800$ users were available, the running time was reduced by $0.82$ seconds from $2,000$ to $2,500$ formal reports. The reduce was $0.32$ seconds with $900$ users. We found that number of users had a more effect on running time per iteration than number of formal reports. For instance, when  $2,000$ formal reports were available, the running time was increased by $1.2$ seconds from $300$ users to $400$ users. However, the increase was only $0.4$ seconds from $2,000$ to $2,500$ formal reports with  $300$ users.
\vspace{-0.2cm}
\begin{table}[!hbp]
\centering
\small
\caption{The relation between running time per iteration (seconds) and number of formal reports and number of users: generally, running time per iteration increased with number of formal reports and number of users; number of users had a more effect on running time per iteration than number of formal reports.}
\vspace{-0.2cm}
\begin{tabular}{p{0.8cm}|c|c|c|c|c}
\hline\hline
\multicolumn{6}{c}{From $100$ users to $500$ users}\\
\hline
\tabincell{c}{Formal \\ \#report} &100 users&200 users&300 users&400 users&500 users\\
\hline
500&1.6225&    1.7930&    2.8969 &   3.5768   & 3.8594\\
\hline
1000&1.4199&    1.7098&    2.6765&    3.4179&    3.3053\\
\hline
1500&1.0185&    1.7127&    2.6551&    3.3280&    4.0004\\
\hline
2000& 1.6106&    1.9634&    3.2525&    4.4560&    4.6050\\
\hline 2500& 1.2893&    2.3403&    3.6309&    4.4778&    4.9070\\
\hline\hline
\multicolumn{6}{c}{From $600$ users to $1000$ users}\\
\hline
\tabincell{c}{Formal \\ \#report} &600 users&700 users&800 users&900 users&1000 users\\
\hline
500&4.0152&    4.1292&    4.8341&    4.8169&    4.8815\\\hline
1000&4.3464&    4.9213&    5.2131&    5.2943&    5.4708\\\hline
1500&4.1058&    3.9091&    5.3806&    5.6140&    5.8158\\\hline
2000&4.6901&    5.4462&    5.9321&    6.0469&    6.2645
\\\hline
2500&5.3191&    5.4397&    5.1165&    5.6264&    5.8152\\\hline\hline
\multicolumn{6}{c}{From $1100$ users to $1500$ users}\\
\hline \tabincell{c}{Formal \\ \#report} &1100 users&1200 users&1300 users&1400 users&1500 users\\
\hline
500&5.7781&    6.2938&    6.8072&    6.9020&    7.3748\\
\hline
1000&6.1258&    7.3894&    7.6492&    7.8697&    7.2863\\
\hline
1500&6.5922&    7.6391&    8.3772&    8.4165&    8.7627\\
\hline
2000&7.1287&    7.8769&    8.3860&    9.1307&    9.1151\\\hline 
2500&7.6745&    8.2669&    8.8098&    9.6669&    9.7889\\\hline\hline
\end{tabular}
\label{tab:running time}
\end{table}
\vspace{-0.6cm}
\subsubsection{Scalability analysis} 
\begin{figure}
\begin{minipage}
{0.49\linewidth}
\centerline{\includegraphics[width=0.9\columnwidth]
{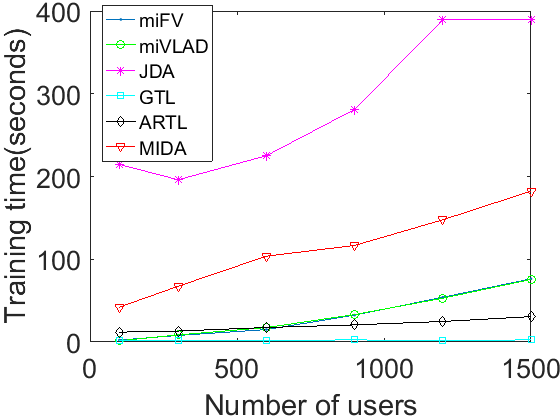}}
\centerline{(a)  Scalability on number} \centerline{of users}
\end{minipage}
\hfill
\begin{minipage}
{0.49\linewidth}
\centerline{\includegraphics[width=0.9\columnwidth]
{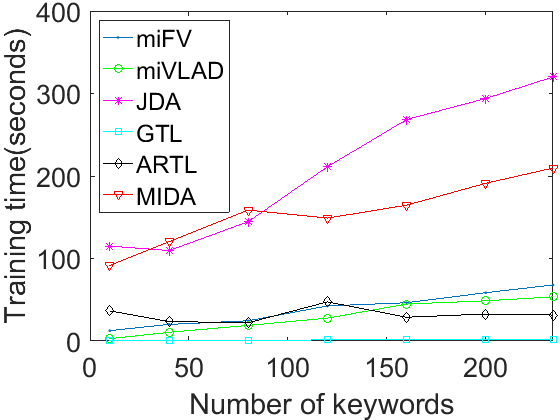}}
\centerline{(b) Scalability on number} \centerline{of keywords}
\end{minipage}
\vspace{-0.3cm}
\caption{Scalability on number of keywords and users: all methods increased linearly with the number of keywords and users.}
\label{fig:scalability}
\vspace{-0.9cm}
\end{figure}

\indent To examine the scalability of the MIDA, we
measured the training times of all methods when varying number of users and number of keywords. The training time was calculated by the average of 5-fold cross validation.\\
\indent Figure \ref{fig:scalability}(a) compares the running time for all methods when the number of users changed from 100 to 1500. Basically, the running time of all
 methods increased linearly with number of users. Among them, the  ARTL and the GTL required the shortest running time compared with other methods. The miFV and the miVLAD were also very efficient even though they were multi-instance methods. Compared with all baselines, the MIDA performs the most work. However, the MIDA effectively reduced computational time by the parallel computing strategy of the ADMM. Surprisingly, the JDA was the slowest method among all baselines. It doubled the training time of the MIDA when $1,500$ users were used for training.\\
\indent To examine the scalability for an increasing number of keywords, Figure \ref{fig:scalability}(b) illustrates the running time of all methods when number of keywords jumped from 10 to 234. Similar to the patterns shown in Figure \ref{fig:scalability}(a), the running time of
all methods increased linearly with number of keywords, which demonstrated that our MIDA was scalable with respect to number of keywords. Note that the ARTL, the GTL, the miFV and the miVLAD increased smoothly with number of keywords. The JDA skyrocketed to $300$ seconds when $234$ keywords were included for training. 
\subsubsection{Case Studies}
\begin{figure}
\begin{minipage}
{0.5\linewidth}
\centerline{\includegraphics[width=0.9\columnwidth]{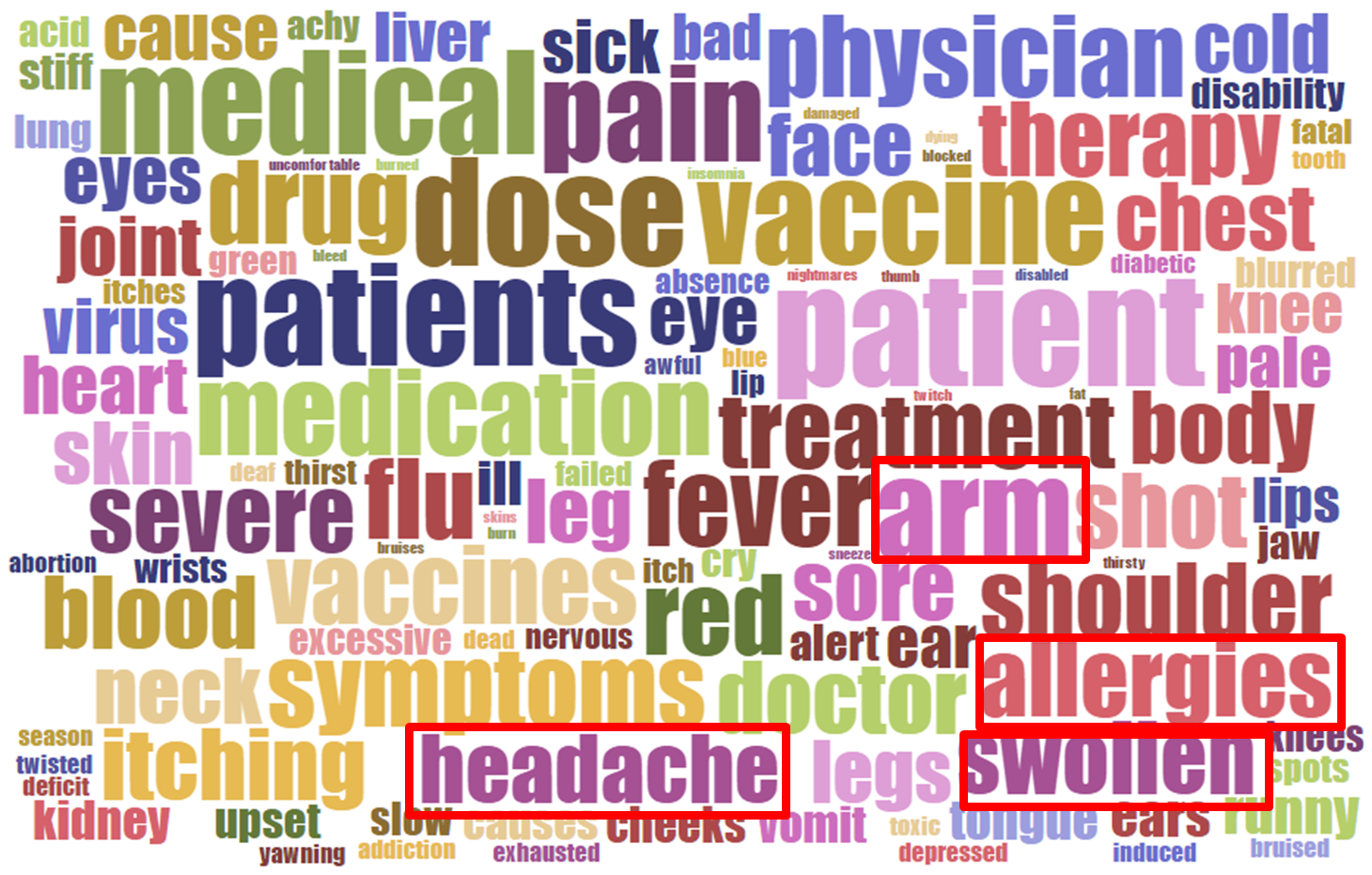}}
\centerline{(a) keyword frequencies of formal reports.}
\end{minipage}
\vfill
\begin{minipage}
{0.49\linewidth}
\centerline{\includegraphics[width=0.9\columnwidth]{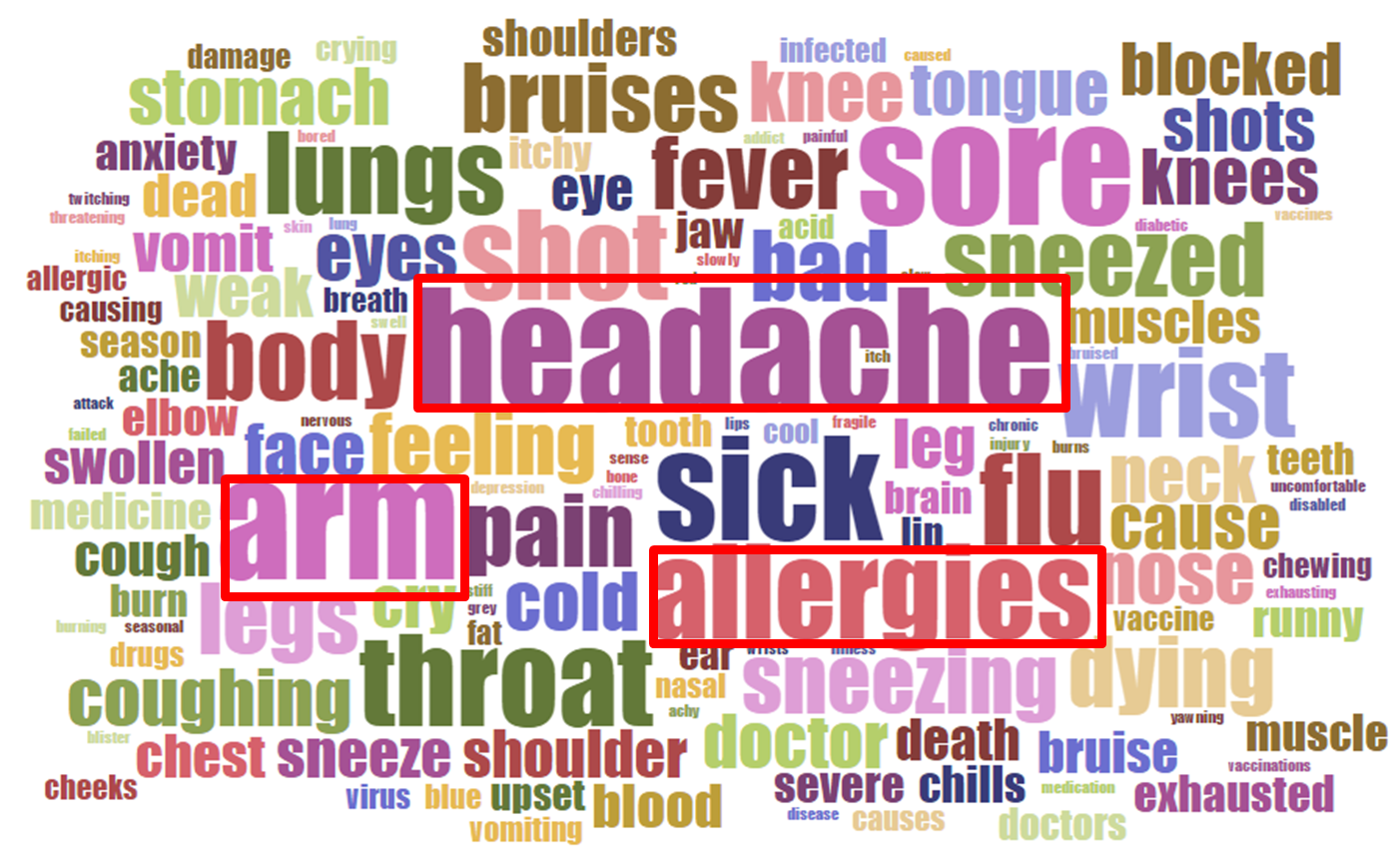}}
\centerline{(b) keyword frequencies of} \centerline{adverse-relevant tweets.}
\end{minipage}
\hfill
\begin{minipage}
{0.49\linewidth}
\centerline{\includegraphics[width=0.9\columnwidth]{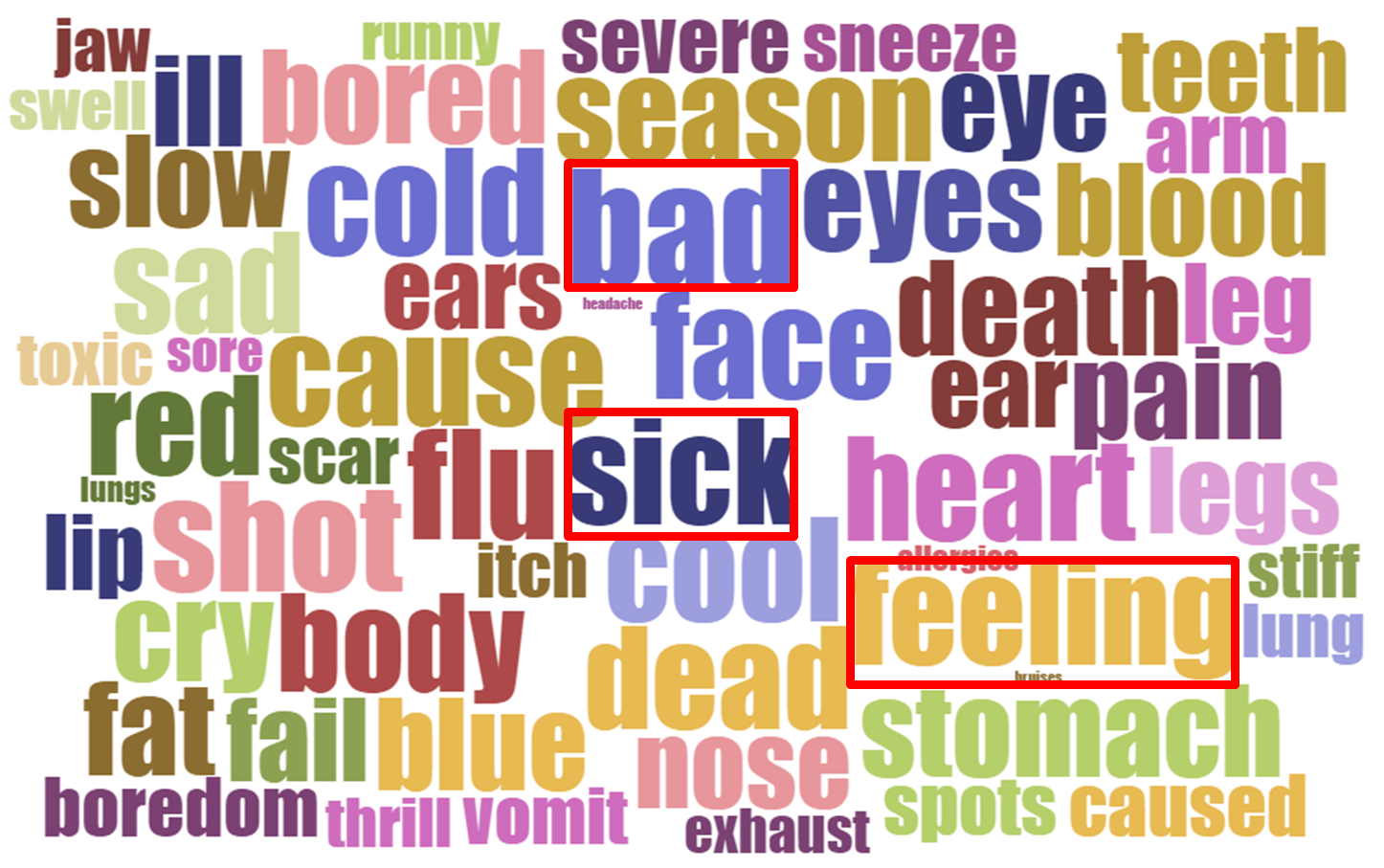}}
\centerline{(c) keyword frequencies of}
\centerline{adverse-irrelevant tweets.}
\end{minipage}
\vspace{-0.3cm}
\caption{Keyword frequencies of formal reports and tweets: extracted adverse-relevant tweets shared a sense of similarity with formal reports, but were different from adverse-irrelevant tweets.}
\label{fig:keyword frequency}
\vspace{-0.6cm}
\end{figure}

\indent We found some benefits from formal reports to improve classifier performance to detect adverse-relevant (i.e., positive) tweets. Figure \ref{fig:keyword frequency} compares the keyword patterns of formal reports, the adverse-relevant tweets and adverse-irrelevant (i.e., negative) tweets identified by our method. In each word cloud, the size of keywords is proportional to their frequencies in the tweet set or the formal report set. In every figure, several important and largest keywords are highlighted in red squares. In Figure \ref{fig:keyword frequency}(a), the largest keywords `physician', `medical', `patients' and `dose' were unique in the formal report domain, but several keywords were also shown to describe vaccine side effects like `headache', `swollen', `arm' and `allergies'. Figure \ref{fig:keyword frequency}(b) showed some largest symptom-descriptive keywords such as `headache', `sore', `arm', `allergies' and `throat'. Among them, keywords `headache', `allergies' and `arm' both appeared in Figure \ref{fig:keyword frequency}(a) and (b), indicating some common symptom descriptions were found in both formal reports and tweets. This implied that MIDA benefited from the adaptation from the formal report domain to the Twitter domain. In Figure \ref{fig:keyword frequency}(c), several largest keywords including `bad', `feeling' and `sick' were general words, which showed that the identified negative tweets were relatively irrelevant to adverse events. The difference of keyword frequencies between Figure \ref{fig:keyword frequency}(b) and (c) and the similarity of that between Figure \ref{fig:keyword frequency}(a) and (b) justified the effectiveness of MIDA as shown in Figure \ref{fig:framework overview}: the similar adverse-relevant tweets are to formal reports, the more distinguishable adverse-relevant tweets are from adverse-irrelevant ones.\\
\indent To further explore the benefits of formal reports to MIDA in a deep insight, Table \ref{tab:case study} illustrates five common symptoms found in both formal reports and tweets extracted by MIDA. Most of them were pain in a certain organ, such as arm pain, shoulder and neck pain and headache. The second and third column displayed formal reports and tweets which described the same symptom. We found that text descriptions were very similar in  formal reports and tweets, which complied with the findings in Figure \ref{fig:keyword frequency}. For example, headache and arm pain were listed as two common symptoms recorded in Table \ref{tab:case study}; correspondingly, keywords `headache' and `arm' were one of the most frequent keywords in Figure \ref{fig:keyword frequency} (a) and (b). This implies that despite word usage differences between formal reports and tweets, they share a sense of similarity in the keyword and description patterns. Therefore, the strategy of MIDA that adapting formal reports to tweets benefits the classifier.
\vspace{-0.4cm}
\begin{table}[!hbp]
\small
\caption{Five common symptom descriptions in both formal reports and  adverse-relevant tweets extracted by MIDA: text descriptions in formal reports and tweets were similar.}
\vspace{-0.2cm}
\begin{tabular}{p{1.2cm}|p{3.3cm}|p{3.3cm}}
\hline \hline
Symptoms &Formal Reports& Adverse-relevant Tweets \\&&Extracted by MIDA\\
\hline
 arm pain&Arm pain for >7 days, sought medical treatment at clinic.&Not only did I fall down the steps, but I got my flu shot and my arm is sore.\\\hline\hline
 shoulder and neck pain&overall aches and pain, but especially in back shoulder blade area and neck.& got my flu shot 30 minutes ago and the SERIOUS ache is spreading  to my shoulder into my neck.\\\hline\hline
 headache&An hour after getting the shot he got a headache and then started throwing up.&I feel that headache slowly coming back after getting shots.\\\hline\hline
 runny nose&Pt received vaccine on 12/11/15.12/14/15 diarrhea, runny nose, cough.&Damn flu shots!  now my nose startin to run.\\\hline\hline
 throat pain&Shortly after patient was vaccinated, she started to feel an itching, tingling feeling in her throat.&Just got a shot! i dnt wanna get sick! I already have a sore throat now.\\\hline
\end{tabular}
\label{tab:case study}
\vspace{-0.6cm}
\end{table}
\section{Conclusion}
\indent Vaccine adverse event detection is a crucial problem
for healthcare. Social media has started to be used to detect adverse events because of its message timeliness and sensor ubiquity. However, it still suffers from prohibitive labeling cost and class imbalance problem, which can be solved by formal reports. In this paper, we propose a novel Multi-instance Domain Adaptation (MIDA) model to minimize the domain differences between formal reports and Twitter data. An efficient algorithm has been developed to
optimize parameters accurately. Various experiments demonstrated that our model was superior to all baselines under six metrics. Case studies showed that similar keyword and description patterns were shown in both formal reports and adverse-relevant tweets.
\label{sec:conclusion}

\bibliographystyle{plain}
\balance
\bibliography{sample-sigconf}

\begin{thebibliography}{10}

\bibitem{Amores2013Multiple}
Jaume Amores.
\newblock Multiple instance classification: Review, taxonomy and comparative
  study.
\newblock {\em Artificial Intelligence}, 201(4):81--105, 2013.

\bibitem{Andrews2002Support}
Stuart Andrews, Ioannis Tsochantaridis, and Thomas Hofmann.
\newblock Support vector machines for multiple-instance learning.
\newblock {\em Advances in Neural Information Processing Systems},
  15(2):561--568, 2002.

\bibitem{beck2009fast}
Amir Beck and Marc Teboulle.
\newblock A fast iterative shrinkage-thresholding algorithm for linear inverse
  problems.
\newblock {\em SIAM journal on imaging sciences}, 2(1):183--202, 2009.

\bibitem{boyd2011distributed}
Stephen Boyd, Neal Parikh, Eric Chu, Borja Peleato, and Jonathan Eckstein.
\newblock Distributed optimization and statistical learning via the alternating
  direction method of multipliers.
\newblock {\em Foundations and Trends{\textregistered} in Machine Learning},
  3(1):1--122, 2011.

\bibitem{chattopadhyay2012multisource}
Rita Chattopadhyay, Qian Sun, Wei Fan, Ian Davidson, Sethuraman Panchanathan,
  and Jieping Ye.
\newblock Multisource domain adaptation and its application to early detection
  of fatigue.
\newblock {\em ACM Transactions on Knowledge Discovery from Data (TKDD)},
  6(4):18, 2012.

\bibitem{Chen2014Flu}
Liangzhe Chen, K.~S. M.~Tozammel Hossain, Patrick Butler, Naren Ramakrishnan,
  and B.~Aditya Prakash.
\newblock Flu gone viral: Syndromic surveillance of flu on twitter using
  temporal topic models.
\newblock In {\em IEEE International Conference on Data Mining}, pages
  755--760, 2014.

\bibitem{daume2009frustratingly}
Hal Daum{\'e}~III.
\newblock Frustratingly easy domain adaptation.
\newblock {\em arXiv preprint arXiv:0907.1815}, 2009.

\bibitem{Dietterich1997Solving}
Thomas~G Dietterich, Richard~H Lathrop, and Tom{\'a}s Lozano-P{\'e}rez.
\newblock Solving the multiple instance problem with axis-parallel rectangles.
\newblock {\em Artificial intelligence}, 89(1-2):31--71, 1997.

\bibitem{duan2012domain}
Lixin Duan, Ivor~W Tsang, and Dong Xu.
\newblock Domain transfer multiple kernel learning.
\newblock {\em IEEE Transactions on Pattern Analysis and Machine Intelligence},
  34(3):465--479, 2012.

\bibitem{duan2012learning}
Lixin Duan, Dong Xu, and Ivor Tsang.
\newblock Learning with augmented features for heterogeneous domain adaptation.
\newblock {\em arXiv preprint arXiv:1206.4660}, 2012.

\bibitem{fleuret2003scale}
Fran{\c{c}}ois Fleuret and Hichem Sahbi.
\newblock Scale-invariance of support vector machines based on the triangular
  kernel.
\newblock In {\em 3rd International Workshop on Statistical and Computational
  Theories of Vision}, pages 1--13, 2003.

\bibitem{fu2009instance}
Zhouyu Fu and Antonio Robles-Kelly.
\newblock An instance selection approach to multiple instance learning.
\newblock In {\em Computer Vision and Pattern Recognition, 2009. CVPR 2009.
  IEEE Conference on}, pages 911--918. IEEE, 2009.

\bibitem{hong2016convergence}
Mingyi Hong, Zhi-Quan Luo, and Meisam Razaviyayn.
\newblock Convergence analysis of alternating direction method of multipliers
  for a family of nonconvex problems.
\newblock {\em SIAM Journal on Optimization}, 26(1):337--364, 2016.

\bibitem{iso2016forecasting}
Hayate Iso, Shoko Wakamiya, and Eiji Aramaki.
\newblock Forecasting word model: Twitter-based influenza surveillance and
  prediction.
\newblock In {\em COLING}, pages 76--86, 2016.

\bibitem{K2013Real}
L.~K.
\newblock Real-time disease surveillance using twitter data: demonstration on
  flu and cancer.
\newblock In {\em ACM SIGKDD International Conference on Knowledge Discovery
  and Data Mining}, pages 1474--1477, 2013.

\bibitem{khan2009survey}
Shehroz~S Khan and Michael~G Madden.
\newblock A survey of recent trends in one class classification.
\newblock In {\em Irish Conference on Artificial Intelligence and Cognitive
  Science}, pages 188--197. Springer, 2009.

\bibitem{kulis2011you}
Brian Kulis, Kate Saenko, and Trevor Darrell.
\newblock What you saw is not what you get: Domain adaptation using asymmetric
  kernel transforms.
\newblock In {\em Computer Vision and Pattern Recognition (CVPR), 2011 IEEE
  Conference on}, pages 1785--1792. IEEE, 2011.

\bibitem{Kumar2016Audio}
Anurag Kumar and Bhiksha Raj.
\newblock Audio event detection using weakly labeled data.
\newblock In {\em Proceedings of the 2016 ACM on Multimedia Conference}, pages
  1038--1047. ACM, 2016.

\bibitem{lamb2013separating}
Alex Lamb, Michael~J Paul, and Mark Dredze.
\newblock Separating fact from fear: Tracking flu infections on twitter.
\newblock In {\em HLT-NAACL}, pages 789--795, 2013.

\bibitem{lampos2010flu}
Vasileios Lampos, Tijl De~Bie, and Nello Cristianini.
\newblock Flu detector-tracking epidemics on twitter.
\newblock In {\em Joint European Conference on Machine Learning and Knowledge
  Discovery in Databases}, pages 599--602. Springer, 2010.

\bibitem{lipp2016variations}
Thomas Lipp and Stephen Boyd.
\newblock Variations and extension of the convex--concave procedure.
\newblock {\em Optimization and Engineering}, 17(2):263--287, 2016.

\bibitem{long2014adaptation}
Mingsheng Long, Jianmin Wang, Guiguang Ding, Sinno~Jialin Pan, and S~Yu Philip.
\newblock Adaptation regularization: A general framework for transfer learning.
\newblock {\em IEEE Transactions on Knowledge and Data Engineering},
  26(5):1076--1089, 2014.

\bibitem{long2014transfer}
Mingsheng Long, Jianmin Wang, Guiguang Ding, Dou Shen, and Qiang Yang.
\newblock Transfer learning with graph co-regularization.
\newblock {\em IEEE Transactions on Knowledge and Data Engineering},
  26(7):1805--1818, 2014.

\bibitem{long2013transfer}
Mingsheng Long, Jianmin Wang, Guiguang Ding, Jiaguang Sun, and Philip~S Yu.
\newblock Transfer feature learning with joint distribution adaptation.
\newblock In {\em Proceedings of the IEEE international conference on computer
  vision}, pages 2200--2207, 2013.

\bibitem{Metke2014Evaluation}
Alejandro Metke-Jimenez, Sarvnaz Karimi, and Cecile Paris.
\newblock Evaluation of text-processing algorithms for adverse drug event
  extraction from social media.
\newblock In {\em International Workshop on Social Media Retrieval and
  Analysis}, pages 15--20, 2014.

\bibitem{pan2010cross}
Sinno~Jialin Pan, Xiaochuan Ni, Jian-Tao Sun, Qiang Yang, and Zheng Chen.
\newblock Cross-domain sentiment classification via spectral feature alignment.
\newblock In {\em Proceedings of the 19th international conference on World
  wide web}, pages 751--760. ACM, 2010.

\bibitem{pan2010survey}
Sinno~Jialin Pan and Qiang Yang.
\newblock A survey on transfer learning.
\newblock {\em IEEE Transactions on knowledge and data engineering},
  22(10):1345--1359, 2010.

\bibitem{paul2011you}
Michael~J Paul and Mark Dredze.
\newblock You are what you tweet: Analyzing twitter for public health.
\newblock {\em Icwsm}, 20:265--272, 2011.

\bibitem{ro1973pattern}
Duda Ro and Hart Pe.
\newblock Pattern classification and scene analysis.
\newblock 1973.

\bibitem{signorini2011use}
Alessio Signorini, Alberto~Maria Segre, and Philip~M Polgreen.
\newblock The use of twitter to track levels of disease activity and public
  concern in the us during the influenza a h1n1 pandemic.
\newblock {\em PloS one}, 6(5):e19467, 2011.

\bibitem{vapnik1998statistical}
Vladimir~Naumovich Vapnik and Vlamimir Vapnik.
\newblock {\em Statistical learning theory}, volume~1.
\newblock Wiley New York, 1998.

\bibitem{wang2016domain}
Ke~Wang, Jiayong Liu, and Daniel Gonz{\'a}lez.
\newblock Domain transfer multi-instance dictionary learning.
\newblock {\em Neural Computing and Applications}, pages 1--10, 2016.

\bibitem{wang2014adaptive}
Qifan Wang, Lingyun Ruan, and Luo Si.
\newblock Adaptive knowledge transfer for multiple instance learning in image
  classification.
\newblock In {\em AAAI}, pages 1334--1340, 2014.

\bibitem{wang2015global}
Yu~Wang, Wotao Yin, and Jinshan Zeng.
\newblock Global convergence of admm in nonconvex nonsmooth optimization.
\newblock {\em arXiv preprint arXiv:1511.06324}, 2015.

\bibitem{Wei2014Scalable}
Xiu~Shen Wei, Jianxin Wu, and Zhi~Hua Zhou.
\newblock Scalable multi-instance learning.
\newblock In {\em IEEE International Conference on Data Mining}, pages
  1037--1042, 2014.

\bibitem{weiss2016survey}
Karl Weiss, Taghi~M Khoshgoftaar, and DingDing Wang.
\newblock A survey of transfer learning.
\newblock {\em Journal of Big Data}, 3(1):9, 2016.

\bibitem{Yomtov2013Postmarket}
E~Yomtov and E~Gabrilovich.
\newblock Postmarket drug surveillance without trial costs: Discovery of
  adverse drug reactions through large-scale analysis of web search queries.
\newblock {\em Journal of Medical Internet Research}, 15(6):e124, 2013.

\bibitem{Yu2013LibShortText}
H~Yu, C~Ho, Y~Juan, and C~Lin.
\newblock Libshorttext: A library for short-text classification and analysis.
\newblock {\em Rapport interne, Department of Computer Science, National Taiwan
  University. Software available at http://www. csie. ntu. edu. tw/\~{}
  cjlin/libshorttext}, 2013.

\bibitem{zhang2009multiple}
Dan Zhang and Luo Si.
\newblock Multiple instance transfer learning.
\newblock In {\em Data Mining Workshops, 2009. ICDMW'09. IEEE International
  Conference on}, pages 406--411. IEEE, 2009.

\bibitem{zhao2016online}
Liang Zhao, Feng Chen, Chang-Tien Lu, and Naren Ramakrishnan.
\newblock Online spatial event forecasting in microblogs.
\newblock {\em ACM Transactions on Spatial Algorithms and Systems (TSAS)},
  2(4):15, 2016.

\bibitem{Zhou2008Multi}
Zhi~Hua Zhou, Yu~Yin Sun, and Yu~Feng Li.
\newblock Multi-instance learning by treating instances as non-i.i.d. samples.
\newblock {\em Computer Science}, pages 1249--1256, 2008.

\bibitem{zhu2011heterogeneous}
Yin Zhu, Yuqiang Chen, Zhongqi Lu, Sinno~Jialin Pan, Gui-Rong Xue, Yong Yu, and
  Qiang Yang.
\newblock Heterogeneous transfer learning for image classification.
\newblock In {\em AAAI}, 2011.

\end{thebibliography}
\end{document}